\documentclass[mathserif]{beamer}

\usepackage{amsmath,amsthm,amssymb}
\usepackage{graphicx}
\usepackage{booktabs} 
\usepackage{verbatim}

\mode<presentation>
\usepackage[english]{babel}
\usepackage[latin1]{inputenc}
\usepackage{times}
\usepackage[T1]{fontenc}

\numberwithin{equation}{section}

\theoremstyle{plain}
\newtheorem{thm}{Theorem}[section]

\theoremstyle{definition}

\theoremstyle{remark}

\newcommand{\tb}{\textbf}

\newcommand{\ra}{\rightarrow}
\newcommand{\noi}{\noindent}

\newcommand{\mb}{\mathbb}
\newcommand{\mr}{\mathrm}

\newcommand{\mc}{\mathcal}
\renewcommand{\t}{\text}
\renewcommand{\it}{\textit}
\renewcommand{\i}{\item}

\renewcommand{\th}{\theta}
\newcommand{\Th}{\Theta}
\renewcommand{\l}{\lambda}

\newcommand{\Sig}{\Sigma}
\newcommand{\sig}{\sigma}
\newcommand{\ep}{\epsilon}

\newcommand{\alp}{\alpha}

\newcommand{\til}{\tilde}

\newcommand{\del}{\delta}
\newcommand{\un}{\underline}
\newcommand{\p}{\pause}

\title{Nonparametric Bayes Classification via Learning of Affine Subspaces}

\author[Abhishek Bhattacharya]{ Abhishek Bhattacharya \\  {\em Indian Statistical Institute}  \\
                               \vspace{5mm}
                                based on the paper \tb{{\em Density Estimation and Classification via Bayesian Nonparametric Learning of Affine Subspaces}} jointly with David Dunson \& Garritt Page, 2012
                              }
\date[2013]{January 10, 2013}

\begin{document}

\begin{frame}
\titlepage
\end{frame}

\begin{frame} 
\frametitle{Contents}
\tableofcontents 
\end{frame}


\section{Motivation \& Goal}

\begin{frame}
\frametitle{What are we interested in?}

\begin{itemize}
\i Build efficient nonparametric Bayes classifiers in presence of many predictors.\p
\vspace{4mm}
\i Different cell probabilities allowed to vary non-parametrically based on a few coordinates expressed as  linear combinations of the predictors. \p
\vspace{4mm}
\i Model parameters clearly interpretable and provide insight to which predictors are important in constructing accurate classification boundaries. \p
\vspace{4mm}
\i Estimated cell probabilities consistent in weak and strong sense.  \p
\vspace{4mm}
\i Data applications support the results. 
\end{itemize}
\end{frame}

\section{Framework}
\begin{frame}
\frametitle{Affine Subspace Characterization}
 \begin{itemize}
  \i Let $S$ be an affine subspace of $\Re^m$ of dimension $k$ ($k \ll m$).
  \vspace{4mm}
  \i Let $\th \in \Re^m$ be the projection of the origin in $S$ and $R \in \Re^{m\times m}$ the projection matrix of the  linear subspace parallel to $S$.
  \vspace{4mm}
  \i Hence $R = R' = R^2$, rank$(R) = k$, $R\th = 0$.
  \vspace{4mm}
  \i Let $R = UU'$, $U \in V_{k,m} = \{ U \in \Re^{m\times k}: \ U'U = I_k \}$ - the Steifel manifold.
  \vspace{4mm}
  \i Any $x \in S$ can be given isometric coordinates $\til x = U'x \in \Re^k$ s.t. $x = U\til x + \th$.
  \end{itemize}

\end{frame}

\begin{frame}
 \begin{itemize}
  \i For $x \in \Re^m$, its projection $P_S(x) = Rx + \th$ has coordinates $U'x \in \Re^k$.
  \vspace{4mm}
  \i The residual $R_S(x) = x - P_S(x)$ lies in a linear subspace $S^\bot$ perpendicular to $S$ with projection matrix $I - R = VV'$, $V \in  V_{m-k,m}$, $V'U = 0$.
  \vspace{4mm}
  \i It has coordinates $V'(x-\th)$ in $\Re^{m-k}$.  
 \end{itemize}

\end{frame}

\section{Model}
\begin{frame}
\frametitle{Joint Density Model}
 \begin{itemize}
  \item Let $X$ denote the predictor in $\Re^m$ and $Y$ a categorical response taking values in $\mb{Y} = \{1,\ldots, c\}$.
  \vspace{4mm}
  \i Will estimate the conditional class probabilities by modeling the joint of $(X, Y)$ s.t. $Y$ depends on $X$ only through its projection onto $S$.
  \vspace{4mm}
  \i $(P_S(X),Y)$ has a nonparametric kernel mixture density in $S \times M_c$ while independently $R_S(X)$ follows a mean zero parametric model on $S^\bot$.
  \end{itemize}

\end{frame}

\begin{frame}
 \begin{itemize}
  \i Say $(U'X, Y) \sim \int_{\Re^k \times S_c} N_k(x; \mu, \Sig_1)M_c(y;\nu) P(d\mu d\nu)$ where 
  \vspace{4mm}
  \i $N_k$ denotes the $k$-variate Normal kernel, 
	\vspace{4mm}
	\i $M_c(y;\nu) = \prod_{l=1}^c \nu_l^{I(y=l)}$ is the multinomial kernel and
     \[
       S_c = \{ \nu \in [0,1]^c: \sum_l \nu_l =1 \}.
     \]
 \vspace{4mm}
  \i Independently $V'(X-\th) \sim N_{m-k}(0, \Sig_2)$.
  \end{itemize}

\end{frame}
  
  \begin{frame}
 \begin{itemize}
  \i Then $(X, Y) \sim \int_{\Re^k \times S_c} N_m(x; U\mu + \th, \Sig)M_c(y;\nu) P(d\mu d\nu)$ where
  \vspace{4mm}
  \i $\Sig =  U\Sig_1U' + V\Sig_2V'$.
  \vspace{4mm}
  \i Wlog can take $\Sig_1$ and $\Sig_2$ to be diagonal.  
  \vspace{4mm}
  \i For sparsity assume $\Sig_2 = \sig_0^2 I_{m-k}$, i.e. the $X$ residuals are homogeneously distributed.
	\vspace{4mm}
	\i Let  $\Sig_1 = \t{diag}(\sig_1^2,\ldots,\sig_k^2)$.
	\end{itemize}

\end{frame}
  
  \begin{frame}
 \begin{itemize}
  \i Then $\Sig =  U(\Sig_1 - \sig_0^2I_k)U' + \sig_0^2I_m$ and the model parameters are
  \vspace{4mm}
  \i $k$, $U \in V_{k,m}$, $\th \in \Re^m$ satisfying $U'\th=0$, $\un{\sig} = (\sig_0,\sig_1,\ldots,\sig_k)$ - a positive vector and $P$ - a probability on $\Re^k \times S_c$.
  \vspace{4mm}
  \i For Bayesian n.p. inference set priors on the parameters s.t. the induced prior on the joint density has full support and the posterior estimate is consistent.
  \end{itemize}
\end{frame}

\section{Prior Choice}

  \begin{frame}
	\frametitle{Prior Choice on $\Th$}
 \begin{itemize}
  \item Common prior choice on $\Th = (k, U, \th, \un{\sig}, P)$ that preserves conjugacy can be 
  \vspace{4mm}
  \i a discrete prior on $k$ and given $k$,  
  \vspace{4mm}
  \i a matrix Bingham-von Mises-Fisher density on $U$ which has the form proportional to $\exp\t{Tr}(UA + UBU'C)$,
	\vspace{4mm}
  \i a $m$-variate Normal on $\th$ restricted to the space of vectors orthogonal to $U$, 
	\vspace{4mm}
	\i inverse-Gamma priors on the elements of $\un{\sig}$, and,
  \end{itemize}
\end{frame}

\begin{frame}
 \begin{itemize}
 \i a Dirichlet process (DP) prior on $P$: $P \sim \t{DP}\left(w_0 (P_0\otimes Q_0)\right)$, where $P_0$ is a $k$-variate Normal
  and $Q_0$ a Dirichlet distribution on $S_c$.
 \vspace{4mm}
 \i When $P$ is discrete, say, $P = \sum_{j=1}^\infty w_j \del_{(\mu_j,\nu_j)}$, then
\[
P(Y=y | X=x;\Th) = \sum_{j=1}^\infty \til w_j(U'x)  M_c(y; \nu_j)
\]
where  $\til w_j(x) = \frac{w_j N_k(x; \mu_j,\Sig_1)}{\sum_{i=1}^\infty w_i N_k(x; \mu_i,\Sig_1)}$, $x\in \Re^k$.   
\vspace{4mm}
\i Markov chain Monte Carlo (MCMC) methods can be employed to draw from the posterior.
\vspace{4mm}
\i Choice of o.n. basis leads to rapid convergence and avoids large dimensional matrix inversion.
\end{itemize}
\end{frame}

\section{Weak Posterior Consistency}

\begin{frame}
\frametitle{Consistency of the Conditional Class Probabilities}
To show that the conditional density of Y given X under the posterior is consistent.\\
\vspace{4mm}
\noi Assume the following on $f_t$ - the true joint density of (X,Y).
  
\begin{enumerate}
\i $0 < f_t(x,y) < A$ for some constant $A$ for all $(x,y) \in \Re^m \times \mb{Y}$.
\i  $E_t|\log\{f_t(X,Y)\}| < \infty$.
\i For some $\del > 0$, $E_t \log\frac{f_t(X,Y)}{f_{\del}(X,Y)} < \infty$, where $f_{\del}(x,y) = \mathop{\inf}_{\til x : \| \til x - x \| <\del} f_t(\til x, y)$.
\i For some $\alp > 0$, $E_t \|X\|^{2(1+\alp)m} < \infty$.\\
\end{enumerate}

Here $E_t$ denotes expectation under $f_t$.
\end{frame}

\begin{frame}
\begin{itemize}
\i Define probability $\til P_t$ on $\Re^m \times S_c$ as
\begin{align*}
 \til P_t(d\mu d\nu) = \sum_{j=1}^c f_t(\mu,j) d(\mu) \del_{e_j}(d\nu)
\end{align*}
where $e_j$ is the vector with $1$ as $j$th coordinate and zeros elsewhere.
\vspace{4mm}

\i Set priors on the parameters such that given $k$; ($U$, $\th$), $\un{\sig}$ and $P$ are conditionally independent.
\vspace{4mm}

\i Let $(\mb{X}_n, \mb{Y}_n) = (X_1,Y_1), \ldots, (X_n,Y_n)$ iid $f_t$.
\end{itemize}
\end{frame}

\begin{frame}
\frametitle{Weak Posterior Consistency (WPC)}
\begin{thm}[Weak Posterior Consistency (WPC)]\label{wpc}
Let $Pr(k=m)>0$ and the conditional priors on $\un{\sig}$ and $P$ given $k=m$ contain $\un{0}$ and $\til P_t$ in their weak supports respectively.  
Then under assumptions {\bf 1}-{\bf 4} on $f_t$, the Kullback-Leibler (KL) condition is satisfied by the induced prior on $f$ at $f_t$.
\end{thm}

\break

The proof runs on the same lines of the proof of Theorem 3.1. {\em Bhattacharya, Page \& Dunson 2012}.\\
\p
\vspace{4mm}

This in turn implies a.s. WPC which implies $\forall \ep>0$,
\[ \Pi_n\left\{ \lvert P(Y=y | X\in U; \Th) - P_t(Y=y | X\in U) \rvert > \ep \right\} \ra 0 \t{ a.s. } P_t \]
where $\Pi_n$ denotes the posterior of $\Th$ given $(\mb{X}_n, \mb{Y}_n)$. 

\end{frame}

\section{Strong Posterior Consistency}

\begin{frame}
\frametitle{Strong Posterior Consistency (SPC)}
\begin{thm}[Strong Posterior Consistency (SPC)]\label{spc}
Assume the conditions for WPC hold. Pick positive constants $a,b,\{\tau_k\}_{k=1}^m$ and $A$ 
and set the prior s.t.
for $k \le m-1$, $\|\th\|^a$ follows a Gamma density, $\max(\un{\sig}) \le A^{1/b}$, and $Pr(\min(\un{\sig}) < n^{-1/b}|k)$ 
decays exponentially with $n$. This holds for e.g. with $\sig_j$s all equal and $\sig_j^{-b}$ following a Gamma density truncated to $[A^{-1},\infty)$.
For the $DP\left(w_k(P_k \otimes Q_0)\right)$ prior on $P$, $k\ge 1$, choose $P_k$ to be a Normal density on $\Re^k$ with variance $\tau_k^2I_k$. Then a.s. SPC results if the constants 
satisfy
$\tau_k^2 > 4A^2$, $a < 2(1+\alp)m$ and $1/a + 1/b <  1/m$.
\end{thm}
 
\end{frame}

\begin{frame}

Proof follows from the proof of Theorem 3.5. {\em Bhattacharya, Page \& Dunson 2012}.\\
\p
\vspace{4mm}

 SPC implies
 
 \begin{align*}
 &\Pi_n\left\{ \int_{\Re^m}\lvert P(Y=y | X=x; \Th) - P_t(Y=y | X=x) \rvert g_t(x)dx > \ep \right\} \\
  &\ra 0 \t{ a.s. } P_t \ \forall \ y
 \end{align*}
 
 with $g_t$ the density of $X$ under $P_t$.
 
 \end{frame}

\begin{frame}
 \begin{itemize} 
\i A Inverse Gamma prior on $\un{\sig}$ satisfies the requirements for weak but not strong posterior consistency.  
\p
\vspace{4mm}
\i In {\em Bhattacharya \& Dunson 2011}, a gamma prior is proved eligible when $k=m$ as long as the hyperparameters are allowed to depend on sample size $n$ in a suitable way.
\vspace{4mm}
\i However there it is assumed that $f_t$ has a compact support. 
\vspace{4mm}
\i The result is expected to hold true in this context too.   
\end{itemize}
\end{frame}

\section{Principal Subspace Classifier}
\begin{frame}
\frametitle{Principal Subspace Classifier (PSC)}
 \begin{itemize} 
 
\i The marginal density of $X$ is
\begin{align*}
X \sim g(x; \Th) & = \int_{\Re^k} N_m(x; \phi(\mu), \Sig) P_1(d\mu), \\
\phi(\mu) = U\mu + \th, \ \Sig & = U\Sigma_1U' + V\Sig_2V',
\end{align*}
$P_1$ is the $\mu$ marginal of $P$.
\vspace{4mm}
\i The $X$ component on which $Y$ depends is the $k$-principal component of $X$ if the eigenvalues of $\Sig_1$ are greater than or equal to those of $\Sig_2$ (and $P$ is non-degenerate).
\vspace{4mm}
\i This holds if $\Sig = \sig_0^2I$.

\end{itemize}
\end{frame}

\begin{frame}
 \begin{itemize} 
 
\i In some sense the model can be considered a Bayesian nonparametric extension of the probabilistic PCA of  {\em Tipping \& Bishop 1999} and {\em Nyamundanda et. al. 2010}. 
\vspace{4mm}
\i The model could also be thought of as a nonparametric extension of the Bayesian Gaussian process latent variable models of {\em Titsias \& Lawrence 2010}
and SVD models of {\em Hoff 2007}.

\end{itemize}
\end{frame}

\section{Estimating the Principal Subspace}
\begin{frame}
\frametitle{Estimating $S$}
\begin{itemize} 
\i To obtain a Bayes estimate for the subspace $S$, choose an appropriate loss function and minimize the Bayes risk w.r.t. the posterior distribution.
\vspace{4mm}
\i $S$ is characterized by its projection matrix $R$ and origin $\th$, i.e. the pair $(R,\th)$.
\vspace{4mm}
\i $R\in \Re^{m\times m}$, $\th \in \Re^m$ satisfy $R = R'= R^2$ and $R\th = 0$. We use $\mc{S}_m$ to denote the space of all such pairs. 
\end{itemize}
\end{frame}

\begin{frame}
\begin{itemize} 
\i One particular loss function on $\mc{S}_m$ is
\begin{equation*} 
L\left((R_1,\th_1), (R_2,\th_2)\right) = \| R_1 - R_2 \|^2 + \| \th_1 - \th_2 \|^2, \ (R_i,\th_i) \in \mc{S}_m,
\end{equation*}
where $\|A\|^2 = \sum_{ij} a_{ij}^2 = \mr{Tr}(AA')$.
\vspace{4mm}
\i Then a point estimate for $(R,\th)$ is the $(R_1,\th_1)$ minimizing the posterior expectation of loss $L$ over $(R_2, \th_2)$, 
provided there is a unique minimizer.
\end{itemize}
\end{frame}

\begin{frame}[label=subspaceestimation]
\begin{thm}[Subspace Estimator]
Let $f(R,\th) = \int_{(R_2,\th_2)}L( (R,\th), (R_2, \th_2) ) dP_n(R_2,\th_2)$, $(R,\th) \in \mc{S}_m$. This function is minimized by
$R = \sum_{j=1}^k U_j U_j'$ and $\th = (I - R)\bar{\th}$ where
$\bar{R}$ and $\bar{\th}$ are the posterior means of $R_2$ and $\th_2$ respectively,
\[
2\bar{R} - \bar{\th}\bar{\th}' = \sum_{j=1}^m \l_j U_j U_j',\ \l_1 \ge \ldots \ge \l_m
\]
is a s.v.d. of $2\bar{R} - \bar{\th}\bar{\th}'$, and $k$ minimizes
$k - \sum_{j=1}^k \l_j$. The minimizer is unique iff there is a unique minimimizer $k$ and $\l_k > \l_{k+1}$ for that $k$.
\end{thm}
\end{frame}

\begin{frame}
\begin{itemize} 
\i Proof follows from {\em Bhattacharya et. al. 2012} and {\em Bhattacharya, A. \& Bhattacharya, R. 2012}.
\p
\vspace{4mm}
\i The relative importance of different features $\{X_1,\ldots,X_m\}$ in explaining $Y$ can then be judged by the magnitude of the corresponding diagonal entry of $R$.
\vspace{4mm}
\i The magnitudes can also be used to group the features according to their relative importance. 
\end{itemize}
\end{frame}

\section{Identifiability of the Principal Subspace}

\begin{frame}
\frametitle{Identifiability of $S$}
\begin{itemize} 

\i $X \sim N_m(0,\Sig)* (P_1\circ\phi^{-1})$, with ``$*$'' denoting convolution.  
\vspace{4mm}
\i The characteristic function of $X$ is
\[
\Phi_X(t) = \exp(-1/2 t'\Sig t) \Phi_{P_1\circ \phi^{-1}}(t), \ t\in \Re^m.
\]
\vspace{3mm}
\i If a discrete $P$ is employed, then $\Sig$ and $P_1\circ\phi^{-1}$ can be uniquely determined from the marginal of $X$.
\vspace{4mm}
\i $P_1\circ\phi^{-1}$ is a distribution on $\Re^m$ supported on $S = \phi(\Re^k)$.

\end{itemize}
\end{frame}

\begin{frame}
\begin{itemize} 

\i Define the \it{affine support} of a probability $Q$, $\mbox{asupp}(Q)$  as the intersection of all affine subspaces having prob. 1.  
It contains the support $\mbox{supp}(Q)$ (but may be larger).  
\vspace{4mm}
\i To identify $S$ and $k$ we assume that $\mbox{asupp}(P_1)$ is $\Re^k$.  
\vspace{4mm}
\i Then asupp$(P_1\circ\phi^{-1})$ is an affine subspace of $\Re^m$ of dimension equal to that of asupp$(P_1)$ $=k$.  

\end{itemize}
\end{frame}

\begin{frame}
\begin{itemize} 

\i Since asupp($P_1\circ\phi^{-1}$) is identifiable, this implies that $k$ is also identifiable as its dimension.  
\vspace{4mm}
\i Since $S$ contains asupp($P\circ\phi^{-1}$) and has dimension equal to that of asupp($P\circ\phi^{-1}$),  $S = \t{asupp}(P\circ\phi^{-1})$.  
\vspace{4mm}
\i Then $R=UU'$ and $\th$ are identifiable as the projection matrix and origin of $S$. 

\end{itemize}
\end{frame}

\section{Illustrations With Real Data Sets}

\begin{frame}
\frametitle{Real Data Examples}
\begin{itemize} 
\i The classifier built (PSC) is used in real data examples and its performance compared with other well known classification methods.
\p
\vspace{5mm}
\i Three such competitors considered are $k$ nearest neighbor (KNN),  mixture discriminant analysis (MDA), and support vector machine (SVM).  
\end{itemize}
\end{frame}

\begin{frame}
\begin{itemize} 
\i KNN is algorithmic based and classifies well in a variety of settings. A range of neighborhood sizes are considered with the one producing the best out of sample prediction ultimately used.    
\p
\vspace{4mm}
\i MDA is a flexible model based Gaussian mixture classifier (see  {\em Hastie \& Tibshirani 1996}).  The number of components in the Gaussian mixture chosen to produce the best out of sample prediction.
\p
\vspace{4mm}
\i SVM is a very accurate classifier and is therefore included.
\p
\vspace{4mm}
\i Out of sample prediction error rates used to compare PSC to the $3$ competitors. 
\end{itemize}
\end{frame}

\subsection{Brain Computer Interface Data}  
\begin{frame}
\frametitle{Brain Computer Interface (BCI) Data} 

\begin{itemize} 
\i The BCI dataset consists of a single person performing 400 trials in each of which he imagined movements with either the left hand or the right hand.
\vspace{4mm}
\i For each trial, EEG recorded from 39 electrodes. 
\vspace{4mm}
\i An autoregressive model of order 3 was fit to each of the resulting 39 time series. 
\end{itemize}
\end{frame}

\begin{frame}
\begin{itemize} 
\i The trial is then represented by the total of $117 = 39 \times 3$ dimensional feature space. 
\vspace{4mm}
\i Goal is to classify each trial as left or right hand movements using the 117 features.   
\vspace{4mm}
\i 200 observations randomly selected to serve as testing data.  
\vspace{4mm}
\i Posterior combinations done with dimension $k$ fixed.
\end{itemize}
\end{frame}

\begin{frame}
\begin{itemize} 
\i To select a $k$ the out of sample prediction error rates and area under the receiver operating characteristic (ROC) curve are employed.
\vspace{4mm}
\i Since low out of sample prediction error rates and large areas under the curve are desirable, a $k$-value at-most $25$ that maximized the difference 
between them is selected.
\vspace{4mm}
\i Following this criteria, $k=3$ chosen.
\vspace{4mm}
\i PSC produces an out of sample prediction error rate of 0.205 compared to 0.51 for KNN, 0.25 for MDA and 0.23 for SVM.  
\end{itemize}
\end{frame}

\subsection{Wisconsin Breast Cancer data set}  
\begin{frame}
\frametitle{Wisconsin Breast Cancer (WBC) data set}  
\begin{itemize}
 \item In this data set the response is breast cancer diagnosis while the covariates consists of 9 nominal variables describing some type of breast tissue cell characteristic. 
 \vspace{4mm}
 \i Although this data set is not high dimensional, it provides a nice illustration of the type of information the PSC can provide regarding associations between covariates and response.    
 \vspace{4mm}
\i Similar to what was done with the BCI data set $k = 3$ is selected.
\vspace{4mm}
\i This results in an out of sample prediction error rate of $0.017$ 
which is smaller than the error rate for KNN (0.035), MDA (0.028) and SVM (0.028).  
\end{itemize}
\end{frame}

\begin{frame}
\begin{itemize}
\i Even though the PSC classifies more accurately than the other methods, what is of particular interest is how each of the 9 tumor attributes influence classification.  
\vspace{4mm}
\i The 9 attributes  (clump thickness, uniformity of cell size, uniformity of cell shape, marginal adhesion, single epithelial cell size, bare nuclei, bland chromatin, normal nucleoli, and mitosis) are all related to a lump being benign or not.  
\vspace{4mm}
\i From the theorem on subspace estimation the estimated principal directions are found in the Table below.   
\end{itemize}
\end{frame}

\againframe{subspaceestimation}

\begin{frame}

\begin{table}[htdp] 
\caption{The $k=3$ principal directions of the Breast Cancer data set along with the row norms}
\begin{center}
\begin{tabular}{l  cccc}
\toprule
Variable & $U_{[,1]}$ & $U_{[,2]}$ & $U_{[,3]}$ & norm \\ \midrule 
clump thickness  &-0.294 & 0.233 &   0.453     &0.588 \\ 
uniformity of cell size &  -0.399 & -0.132 &  -0.189    &0.460\\ 
uniformity of cell shape & -0.395 & -0.102 &  0.0172   &0.408\\
marginal adhesion & -0.314 & -0.007 &  -0.477    &0.571 \\
single epithelial cell size & -0.231 & -0.181 &  -0.307 &0.424   \\
bare nuclei & -0.450 & 0.713 &   0.101     &0.849 \\
bland chromatin & -0.295 & -0.032 &  -0.194    &0.354 \\
normal nucleoli &-0.376 & -0.587 &  0.543     &0.883   \\
mitosis &-0.121 & -0.173 &  -0.305    &0.371 \\
\bottomrule
\end{tabular}
\end{center}
\label{principaldirections}
\end{table}

\end{frame}

\begin{frame}
\begin{itemize}
 \i A way to assess the relative importance of each variable and also provide a means of grouping the variables is to calculate the norm associated with each row of $U$ 
 (i.e. the norm of the corresponding diagonal entry of $R = UU'$). 
\vspace{4mm}
 \i These values can be found under the header ``norm'' in the Table.  
\vspace{4mm}
\i It appears that a bare nuclei and normal nucleoli form a group.  
\vspace{4mm}
\i Another is formed by clump thickness and marginal adhesion.  
\vspace{4mm}
\i Finally it appears that uniformity of cell size, uniformity of cell shape and single epithelial cell size form a group.  
\end{itemize}
\end{frame}

\section{Summary}
\begin{frame}
\frametitle{Summary}
\begin{itemize}
\i A flexible nonparametric model proposed for classification via feature space dimension reduction.
\p
\vspace{4mm}
\i  The model satisfies large support \& consistency conditions.
\p
\vspace{4mm}

\i A simple Gibbs sampler can be implemented with conjugate sampling steps for posterior sampling.
\p
\vspace{4mm}

\i Better performance than commonly used machine learning, computer science and parametric statistical methods.
\end{itemize}
\end{frame}

\begin{frame}
\begin{itemize}
\i These methods are algorithmic or highly parameterized black boxes and apart from classification, provide no further
information specific to the problem being studied.
\p
\vspace{4mm}
\i In addition to building efficient classifiers, the proposed methodology provides insight regarding predictors that are influential in explaining the response - an information applied scientists often highly value.
\p
\vspace{4mm}
\i Can easily be extended to other regression setup. 
\end{itemize}
\end{frame}

\section{Further Work possible}	
\begin{frame}
\frametitle{Further Work possible}	
\begin{itemize}
\i Change the joint kernel choice to build better classifier.
\p
\vspace{4mm}
\i Change the notion of inner product to use non linear predictor transformations to explain the response.
\p
\vspace{4mm}
\i A nonparametric model may be fit on the non-signal predictors as well.
\p
\vspace{4mm}
\i Use other priors besides Dirichlet Process.
\p
\vspace{4mm}
\i Extend to nonparametric hypothesis testing on the lines of {\em Bhattacharya \& Dunson 2012}.
\end{itemize}
\end{frame}

\end{document}